# STruD: Truss Decomposition of Simplicial Complexes

Giulia Preti
ISI Foundation
Turin, Italy
giulia.preti@isi.it

Gianmarco De Francisci Morales
ISI Foundation
Turin, Italy
gdfm@acm.org

Francesco Bonchi
ISI Foundation, Italy
Eurecat, Spain
francesco.bonchi@isi.it

## ABSTRACT

A simplicial complex is a generalization of a graph: a collection of $n$-ary relationships (instead of binary as the edges of a graph), named simplices. In this paper, we develop a new tool to study the structure of simplicial complexes: we generalize the graph notion of truss decomposition to complexes, and show that this more powerful representation gives rise to different properties compared to the graph-based one. This power, however, comes with important computational challenges derived from the combinatorial explosion caused by the downward closure property of complexes.

Drawing upon ideas from itemset mining and similarity search, we design a memory-aware algorithm, dubbed STruD, which is able to efficiently compute the truss decomposition of a simplicial complex. STruD adapts its behavior to the amount of available memory by storing intermediate data in a compact way. We then devise a variant that computes directly the $n$ simplices of maximum trussness. By applying STruD to several datasets, we prove its scalability, and provide an analysis of their structure.

Finally, we show that the truss decomposition can be seen as a *filtration*, and as such it can be used to study the *persistent homology* of a dataset, a method for computing topological features at different spatial resolutions, prominent in Topological Data Analysis.

**ACM Reference Format:**
Giulia Preti, Gianmarco De Francisci Morales, and Francesco Bonchi. 2021. STruD: Truss Decomposition of Simplicial Complexes. In *Proceedings of the Web Conference 2021 (WWW '21), April 19–23, 2021, Ljubljana, Slovenia*. ACM, New York, NY, USA, 11 pages. https://doi.org/10.1145/3442381.3450073

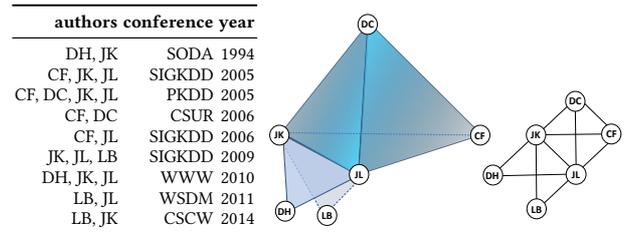

| authors | conference | year |
|---|---|---|
| DH, JK | SODA | 1994 |
| CF, JK, JL | SIGKDD | 2005 |
| CF, DC, JK, JL | PKDD | 2005 |
| CF, DC | CSUR | 2006 |
| CF, JL | SIGKDD | 2006 |
| JK, JL, LB | SIGKDD | 2009 |
| DH, JK, JL | WWW | 2010 |
| LB, JL | WSDM | 2011 |
| LB, JK | CSCW | 2014 |

**Figure 1: Authors, conference, and year of publication of a set of papers (left), the corresponding co-authorship relations represented as a simplicial complex $\mathcal{K}$ (center), and as a simple collaboration network $G$ (right).**

## 1 INTRODUCTION

Graphs have been widely used to model complex relationships (*edges*) between pairs of entities (*vertices*). Aiming to discover useful patterns such as communities, and identify important vertices in a graph, several metrics and substructures have been defined, and applied in settings such as biology, sociology, and Internet topology. Among them, truss decomposition has received considerable attention [21, 40] thanks to its efficiency and effectiveness. Trusses are cohesive subgraphs that are rich in triangles: a relaxation of cliques that can be computed exactly in polynomial time.

However, many real-world interactions occur among more than two entities at once [8, 41]. For example, people are likely to collaborate in groups when writing papers, and often send emails that have multiple recipients. Simple graphs are not able to capture such higher-order relationships, because they cannot distinguish the case of three authors writing three papers in pairs from the case where all three authors collaborate on a single paper. *Hypergraphs* [10] and *simplicial complexes* [2] are higher-order generalizations of simple graphs that can characterize interactions between any number of entities [38]. Hypergraphs generalize graphs by allowing an edge to connect several vertices. A hypergraph $H$ is a pair $(V_H, E_H)$ where $V_H$ is a set of vertices and $E_H$ a family of subsets of vertices called *hyperedges*. Conversely, a simplicial complex is a collection of polytopes such as triangles and tetrahedra, which are called *simplices*. Both these structures can be used to represent any higher-order relation [37]. The key difference between hypergraphs and simplicial complexes is that the latter satisfy the *downward closure* property: every substructure (also known as *face*) of a simplex that is contained in a complex $\mathcal{K}$ is also in $\mathcal{K}$. While it might appear constraining, this property naturally arises in all systems characterized by interactions that are "maximal" [42]: e.g., in scientific collaborations (the authors of a paper are all co-authors) or gene activation pathways (largest group of collectively activated genes).

EXAMPLE 1. *The table in Figure 1 (left) reports authors, conference, and year of publication of a set of papers extracted from DBLP. A typical representation of this type of data is a collaboration network built around the co-authorship relation: two authors (vertices) are connected by an edge if they have been co-authors in some paper. This type of simple graph is reported in Figure 1 (right).*

*A richer representation can be obtained by interpreting each subset of authors of a paper as co-authors, and representing it as a simplex: this leads to the simplicial complex $\mathcal{K}$ illustrated in Figure 1 (center).*

For example, D. Chakrabarti (DC), C. Faloutsos (CF), J. Kleinberg (JK), J. Leskovec (JL) form a tetrahedron because they wrote together a paper at PKDD 2005. Although in the table there is no paper authored solely by (DC, JK, JL), such a triangle is nevertheless part of the simplicial complex $\mathcal{K}$, which represents the fact that these three scientists have been co-authors in a paper (even though with other co-authors).

The simplicial complex conveys information that is lost if the co-authorship relations are modeled as a simple collaboration network. For instance, the latter would represent the case where JK, JL, and DH wrote a paper together in the same way as the case where they wrote papers only in pairs.

Simplices and simplicial complexes have been successfully applied to analyze the organization of the brain [28, 30], understand the mechanics of social contagion [16], predict the appearance of new links [3], and study protein interaction networks [11].

In this paper, we develop a new tool to study the structure of simplicial complexes, by generalizing the graph notion of truss decomposition. The definition of truss decomposition in a graph is based on triangles, i.e., a $k$-truss is a subgraph whose edges participate in (are supported by) at least $k$ triangles.[1] When the input is a simplicial complex, a truss can as well be determined by the existence of higher-order structures.

**Challenges and contributions.** We bridge two different disciplines, i.e., graph mining and Topological Data Analysis (TDA), by generalizing the notion of truss decomposition to simplicial complexes. Our problem statement is a proper generalization of truss decomposition on graphs: if the simplicial complex in input is a collection of only 1-dimensional simplices (edges), or in other terms, it is a graph, then the truss decomposition of the simplicial complex corresponds to that of the graph.

We show several properties of simplicial truss decomposition, and show how this more expressive representation differentiates itself from the graph-based one. However, computing it is much harder than for a graph. Indeed, due to the downward closure property of simplicial complexes that produces a combinatorial explosion of the candidate substructures, the computation becomes especially demanding in terms of memory.

To tackle these significant computation challenges, this paper introduces STruD, an efficient and scalable algorithm for simplicial truss decomposition. A key observation towards developing our method is an *a-priori property* similar to the one from frequent itemset mining: a $(q + 1)$-simplex has simplicial trussness no larger than a $q$-simplex that it contains. Based on this key observation, we derive lower and upper bounds for the simplicial trussness of a simplex which are at the basis of our method. Our algorithm relies on finding *joists*, higher-order generalization of triangles that compose the support of a simplex. The identification of the joists of a simplex is a major computational bottleneck, as their number can be extremely large. To tackle this challenge, we leverage a compact inverted index, and devise a memory-aware strategy that switches to out-of-core operations depending on the memory available. Our extensive empirical assessment confirms the scalability of STruD and of the relevance of the simplicial truss decomposition.

Finally, we show how the simplicial truss decomposition can be interpreted and used as a *filtration* in the context of *persistent homology* [14], a technique used to quantify the shape of the data via summarization of its topological features. Persistent homology requires to define a nested sequence of simplicial complexes called filtration, and tracks the topological features which exhibit long persistence through the filtration. Persistent homology has been applied, among others, to characterize financial markets [39], study multivariate time series [35], and analyze sensor networks [9].

Table 1: Simplicial trussness $tr_\mathcal{K}$ of the simplices in $\mathcal{K}$ in Figure 1 (center), number of joists $|J|$ containing each simplex, and graph trussness $tr_G$ of the edges in $G$ in Figure 1 (right).

| $\sigma^{(q)}$ | $|J|$ | $tr_\mathcal{K}$ | $tr_G$ | $\sigma^{(q)}$ | $|J|$ | $tr_\mathcal{K}$ | $tr_G$ |
|---|---|---|---|---|---|---|---|
| [CF, DC] | 2 | 2 | 2 | [DH, JL] | 1 | 1 | 1 |
| [CF, JK] | 2 | 2 | 2 | [DH, JK, JL] | 0 | 0 | - |
| [CF, JL] | 2 | 2 | 2 | [JK, JL, LB] | 0 | 0 | - |
| [DC, JK] | 2 | 2 | 2 | [DC, JK, JL] | 1 | 1 | - |
| [DC, JL] | 2 | 2 | 2 | [CF, JK, JL] | 1 | 1 | - |
| [JK, JL] | 4 | 2 | 2 | [CF, DC, JL] | 1 | 1 | - |
| [LB, JK] | 1 | 1 | 1 | [CF, DC, JK] | 1 | 1 | - |
| [LB, JL] | 1 | 1 | 1 | [CF, DC, JK, JL] | 0 | - | |
| [DH, JK] | 1 | 1 | 1 | | | | |

The contributions of our work can be summarized as follows:

- We define the truss decomposition of a simplicial complex.
- We design an efficient and scalable algorithm that leverages bounds and techniques from itemset mining and similarity search to compute the simplicial truss decomposition.
- We provide extensive empirical evidence of the scalability of the algorithm, and of the relevance of the simplicial truss decomposition.
- We showcase the use of the simplicial truss decomposition as a filtration to compute a persistent homology.

## 2 PROBLEM DEFINITION

We next provide the formal definitions of the needed notions and of the computational problem addressed in this paper.

Let $V$ be a ground set of elements. A $q$-dimensional simplex, or simply $q$-simplex, is a relation connecting $q + 1$ elements of $V$.

DEFINITION 1 ($q$-SIMPLEX, FACE, COFACE). *A $q$-simplex $\sigma^{(q)}$ is a set of $q + 1$ elements (or vertices) $\sigma^{(q)} = [v_0, \ldots, v_q] \subset V$. Each $r$-simplex $\sigma^{(r)} \subset \sigma^{(q)}$ (with $r < q$) is called a* face *of $\sigma^{(q)}$, and it is called a* coface *when $r = q - 1$.*

According to this definition, a vertex of a graph is a 0-simplex, and an edge a 1-simplex, while a 2-simplex can be seen as a triangle, a 3-simplex a tetrahedron, and so on. For instance, in Figure 1 (center), the triplet of authors (LB, JL, JK) constitutes a 2-simplex $\sigma$, and each pair of authors is a coface of $\sigma$.

DEFINITION 2 (SIMPLICIAL COMPLEX, DIMENSION, $n$-SKELETON). *A simplicial complex $\mathcal{K}$ is a set of simplices such that all the faces of all the simplices in $\mathcal{K}$ are also in $\mathcal{K}$. The dimension of $\mathcal{K}$ is the dimension of its largest simplex, and the n-skeleton of $\mathcal{K}$ is the subset of simplices of $\mathcal{K}$ of dimensions $q \leq n$.*

According to this definition, the 1-skeleton of a simplicial complex corresponds to the underlying (undirected simple) graph of $\mathcal{K}$. For instance, the 1-skeleton of the simplicial complex in Figure 1 (center) is the graph in Figure 1 (right).

---

[1]The original definition of $k$-truss though, requires that each edge take part to at least $k - 2$ triangles, so that a $k$-clique is a $k$-truss [7]. Given that the two definitions are analogous, for convenience in this work we adopt the one requiring $k$ triangles.

In a simple graph $G = (V, E)$, a $k$-truss is defined as the subgraph $G_{T_k}$ of $G$ induced by the edges $S$ that participate in at least $k$ triangles in $S$ [7].[1] According to this definition, for each edge $e$ in the $k$-truss $G_{T_k}$, there exist $k$ pairs of edges $[e_1, e_2]$ such that the edges $[e, e_1, e_2]$ forms a triangle in $G_{T_k}$; as such, a $k$-clique is a $(k-2)$-truss. As an example, the graph in Figure 1 (right) contains the 2-truss formed by the subgraph induced by the vertices (CF, DC, JK, JL), and the 1-truss formed by the whole graph. We generalize this notion to simplicial complexes by considering $q$-simplices instead of edges, and *joists* (defined next) instead of triangles.

DEFINITION 3 (JOIST). *The joist of a simplex $\sigma^{(q)}$ is the set $J_{\sigma^{(q)}}$ of all its cofaces (of which there are $q + 1$).*

DEFINITION 4 (SIMPLICIAL $k$-TRUSS). *Given a simplicial complex $\mathcal{K}$, a $k$-truss in $\mathcal{K}$ is a maximal set of simplices $T_k \subseteq \mathcal{K}$ of dimension greater than 0, such that for each $\sigma^{(q)} = [v_0, \ldots, v_q] \in T_k$ there exist at least $k$ joists $J_1, \ldots, J_k$ such that $\forall i \in [1, k]$, $\sigma^{(q)} \in J_i$ and $\forall \tau \in J_i$, $\tau \in T_k$. The maximal $k$ such that $\sigma^{(q)} \in T_k$ is called* simplicial trussness *of $\sigma^{(q)}$ and denoted as $tr_{\mathcal{K}}(\sigma^{(q)})$.*

EXAMPLE 2. *The simplicial complex in Figure 1 (center) contains a 1-truss and a 2-truss. The 1-truss includes the simplices in Table 1 with simplicial trussness $tr_{\mathcal{K}}$ greater or equal to 1, while the 2-truss includes the simplices with simplicial trussness greater or equal to 2. Differently from the standard truss decomposition, in this case the 1-truss $T_1$ contains also all the 2-simplices (triangles) spanned by the vertices CF, DC, JK, and JL.*

We next provide two key properties of simplicial $k$-trusses. The proofs can be found in Appendix B.

PROPERTY 1 (UNIQUENESS). *The $k$-truss of a simplicial complex $\mathcal{K}$ is unique.*

PROPERTY 2 (CONTAINMENT). *The $(k+1)$-truss of a simplicial complex $\mathcal{K}$ is a subset of its $k$-truss.*

Thanks to the properties of uniqueness and containment we can define a *simplicial truss decomposition*, i.e., the problem of computing all the non-empty $k$-trusses of a simplicial complex $\mathcal{K}$.

PROBLEM 1 (SIMPLICIAL TRUSS DECOMPOSITION). *Given a simplicial complex $\mathcal{K}$, find the simplicial truss decomposition of $\mathcal{K}$, i.e., the sequence of $k$-trusses $\mathcal{T} = [T_1, \ldots, T_K]$, where $K$ is the maximal integer such that $T_K \neq \emptyset$.*

OBSERVATION 1. *Our problem statement is a proper generalization of the standard truss decomposition on graphs. When a simplicial complex $\mathcal{K}$ contains only binary relationships, there exists a bijection $f : \mathcal{T} \mapsto \mathcal{G}_{\mathcal{T}}$ between the simplicial truss decomposition $\mathcal{T}$ of $\mathcal{K}$ and the standard truss decomposition $\mathcal{G}_{\mathcal{T}}$ of the 1-skeleton of $\mathcal{K}$. The bijection maps each $k$-truss $T_k \in \mathcal{T}$ to the $k$-truss $G_{T_k} \in \mathcal{G}_{\mathcal{T}}$ by associating each 1-simplex $[v_1, v_2]$ to the edge $(v_1, v_2)$.*

*For instance, the simplicial truss decomposition of the graph in Figure 1 (right) is $T_1 = \{[CF, DC], [CF, JK], [CF, JL], [DC, JK], [DC, JL], [JK, JL], [LB, JK], [LB, JL], [DH, JL], [DH, JK]\}$ and $T_2 = \{[CF, DC], [CF, JK], [CF, JL], [DC, JK], [DC, JL], [JK, JL]\}$, which maps to the 1-truss $G_{T_1} = G$ and the 2-truss $G_{T_2}$ with edge set $\{(CF, DC), (CF, JK), (CF, JL), (DC, JK), (DC, JL), (JK, JL)\}$.*

---
[1] The original definition of $k$-truss requires each edge to take part to at least $k - 2$ triangles, so that a $k$-clique is a $k$-truss. Given that the two definitions are analogous, for convenience in this work we adopt the one requiring $k$ triangles.

## 3 SIMPLICIAL TRUSS DECOMPOSITION

In this section we present our algorithms for computing the simplicial truss decomposition. A key property (proof in in Appendix B) derives directly from the downward closure of simplicial complexes, and helps greatly to prune the search space, similarly to the *a-priori* property of frequency in *frequent itemset mining* [1].

PROPERTY 3 (A-PRIORI PROPERTY). *A $q$-simplex $\sigma^{(q)}$ that is a face of a $(q+1)$-simplex $\sigma^{(q+1)}$ has trussness $tr_{\mathcal{K}}(\sigma^{(q)}) \geq tr_{\mathcal{K}}(\sigma^{(q+1)})$.*

A corollary of Property 3 is that a $k$-truss $T_k$ is a simplicial complex, because if a $q$-simplex belongs to $T_k$, then all its faces belong to $T_k$ as well. In addition, we can identify a lower bound of the simplicial trussness of any $\sigma^{(q)}$ in $\mathcal{K}$ by looking at the largest simplex that contains $\sigma^{(q)}$ (i.e., that $\sigma^{(q)}$ is a face of):

PROPERTY 4 (LOWER BOUND). *Let $\sigma^{(q)}$ be a $q$-simplex and $\sigma^{(q+h)}$ be the largest simplex in $\mathcal{K}$ such that $\sigma^{(q)} \subset \sigma^{(q+h)}$. Then,*

$$tr_{\mathcal{K}}(\sigma^{(q)}) \geq h - q.$$

Finally, since a simplicial truss is defined by a set of joists, an upper bound of the trussness of $\sigma^{(q)}$ is given by the total number of joists containing $\sigma^{(q)}$:

PROPERTY 5 (UPPER BOUND). *Let $\sigma^{(q)}$ be a $q$-simplex in $\mathcal{K}$ and $J_{\sigma^{(q+1)}}$ indicate a joist of a $(q+1)$-simplex. Then,*

$$tr_{\mathcal{K}}(\sigma^{(q)}) \leq \left|\{J_{\sigma^{(q+1)}} \mid J_{\sigma^{(q+1)}} \subseteq \mathcal{K} \land \sigma^{(q)} \in J_{\sigma^{(q+1)}}\}\right|.$$

We propose algorithms to solve two different variants of the truss decomposition problem. The first algorithm performs the complete truss decomposition of the simplicial complex, and the second one finds the top-$n$ simplices with maximum trussness and given size $q$. Both algorithms follow a 3-step, apriori-like approach that materializes and examines simplices of increasing dimension. In the first step, they extend the simplices retained in the previous iteration by adding an additional vertex; in the second step, they search for all the sets of simplices that form a joist; and in the last step, they compute the trussness of each simplex. The simplices with positive trussness will be extended in the following iteration. The pseudocode of our STRuD algorithm is reported in Algorithm 1.

Recall that a $q$-simplex $\sigma^{(q)}$ belongs to a joist when, together with other $(q + 1)$ $q$-simplices, it forms the cofaces of the same $(q + 1)$-simplex. Therefore, $\sigma^{(q)}$ shares exactly $q$ vertices with any other simplex in the joist. As a consequence, we can run STRuD separately on each connected component of the 1-skeleton of the simplicial complex. Moreover, thanks to Property 3, only the simplices with trussness greater than 0 are extended by the procedure EXTENDSIMPLICES (line 6). This procedure receives in input a set of simplices $S$ and the set of simplices in the connected component under examination $C$, and extends each $\sigma \in S$ by appending each vertex $v \in V_\sigma$, the set of all the vertices that appear in some $\sigma' \in \mathcal{K}$ such that $\sigma \subset \sigma'$. Generating the extensions by using only those vertices guarantees that we create only simplices that exist in $\mathcal{K}$.

Once the set of extended simplices $E$ is computed, we need to find all the joists formed by simplices in $E$. To do so, we need to find, for each $q$-simplex $\sigma \in E$, all the $q$-simplices in $E$ that could be part of a joist with $\sigma$. Therefore, as the size of $E$ increases, the memory required to store the candidate joists increases as well. To deal with

## Algorithm 1 STRUD

**Require:** Simplicial complex $\mathcal{K}$; Max size $d$
**Ensure:** Trussness of the simplices in $\mathcal{K}$
1: $tr \leftarrow \varnothing$
2: **for** $C \in \leftarrow$ CONNECTEDCOMPONENTS($\mathcal{K}$) **do**
3:    $S \leftarrow \varnothing; d \leftarrow \min(d, \max_{\sigma \in C} |\sigma|)$
4:    **for** $q \leftarrow 2$ to $d$ **do**
5:      **if** $q > 2$ **and** $S = \varnothing$ **then break**
6:      $E \leftarrow$ EXTENDSIMPLICES($S, C, q$)
7:      **if** $E = \varnothing$ **then break**
8:      $lb[\sigma] \leftarrow \max_{\tau \in C \wedge \sigma \subseteq \tau}(|\tau|) - |\sigma|$ **for** $\sigma \in E$
9:      $CJ \leftarrow$ FINDJOISTS($E$)
10:      $tr[\sigma] \leftarrow |CJ[\sigma]|$ **for** $\sigma \in E$
11:      **if** $\nexists \sigma$ s.t. $lb[\sigma] \neq tr[\sigma]$ **then**
12:        $S \leftarrow \{\sigma \mid \sigma \in E \wedge tr[\sigma] > 0\}$
13:        **continue**
14:      $S \leftarrow \varnothing$    ▷ simplices to extend in the next iteration
15:      $Q \leftarrow \varnothing$    ▷ ordered queue of simplices to examine
16:      $Q[tr[\sigma]] \leftarrow Q[tr[\sigma]] \cup \{\sigma\}$ **for** $\sigma \in E$
17:      **while** $Q \neq \varnothing$ **do**
18:        $M \leftarrow$ simplices with minimum trussness in $Q$
19:        **for** $\sigma \in M$ **do**
20:          **for** subset $\xi$ of $\sigma$ of size $|\sigma| - 1$ **do**
21:            **for** $v \in CJ[\sigma]$ **do**
22:              $\tau \leftarrow \xi \cup \{v\}$
23:              **if** $tr[\tau] > tr[\sigma]$ **then**
24:                remove $\sigma \setminus \tau$ from $CJ[\tau]$
25:                update $Q$ **if** $tr[\tau]$ has changed
26:          **if** $tr[\sigma] > 0$ **then** $S \leftarrow S \cup \{\sigma\}$
27:          remove $\sigma$ from $Q$
28:      remove from $tr$ each $\sigma$ with $tr[\sigma] = lb[\sigma]$
29: **return** $tr$

## Algorithm 2 FINDJOISTS

**Require:** Set of simplices $E$
**Ensure:** The joists formed by simplices in $E$
1: $CJ \leftarrow \varnothing; I \leftarrow \varnothing$
2: **for** $s \in E$ **do**
3:    $CJ \leftarrow$ MERGE($CJ$, FINDMATCHES($s, I$), $s$)
4:    **for** subset $\xi$ of $\sigma$ of size $|\sigma| - 1$ **do**
5:      $I[\xi] \leftarrow I[\xi] \cup \{s\}$
6: $CJ \leftarrow$ VALIDATEJOISTS($CJ$)
7: **return** $CJ$

8: **function** FINDMATCHES($\sigma, I$)
9:    $CJ \leftarrow \varnothing$
10:    **for** subset $\xi$ of $\sigma$ of size $|\sigma| - 1$ **do**
11:      **for** $\tau \in I[\xi]$ **do** $CJ \leftarrow CJ \cup \{\tau\}$
12:    **return** $CJ$

13: **function** MERGE($CJ, matches, \sigma$)
14:    **for** $\tau \in matches$ **do**
15:      $v \leftarrow \sigma \setminus \tau; u \leftarrow \tau \setminus \sigma$
16:      add $\tau$ to $CJ[\sigma]$ **if** $u > \sigma[|\sigma| - 1]$
17:      add $\sigma$ to $CJ[\tau]$ **if** $v > \tau[|\tau| - 1]$
18:    **return** $CJ$

19: **function** VALIDATEJOISTS($CJ$)
20:    $\mathcal{J} \leftarrow \varnothing$
21:    **for** $\sigma \in CJ$ **do**
22:      $W \leftarrow$ vertices in the simplices in $CJ[\sigma]$ but not in $\sigma$
23:      **for** $w \in W$ **do**
24:        $CJ_\sigma^w \leftarrow \{\tau \mid \tau \in CJ[\sigma] \wedge w \in \tau\}$
25:        **if** $|CJ_\sigma^w| = |\sigma|$ **then**
26:          add $w$ to $\mathcal{J}[\sigma]$
27:          add $\sigma \setminus \tau$ to $\mathcal{J}[\tau]$ **for** $\tau \in CJ_\sigma^w$
28:    **return** $\mathcal{J}$

the large memory requirement, and hence allow the algorithm to be executed also on less powerful machines, we propose two different strategies to find and validate all the candidate joists. The first one is an in-memory strategy, while the second one temporarily stores the candidates in chunks on disk, and then reads one chunk at a time to validate the candidates in it. To decide which strategy to use, we keep track of the memory consumption of the data structures that store the candidates. When we reach the memory limit, we create $M$ chunks, where the $i^{th}$ chunk contains the pairs ($idx_1, idx_2$) such that $idx_1 \mod M = i$ and $idx_2$ is the id of the simplex that could be part of a joist with the simplex with id $idx_1$. Then, if the chunk can be loaded into main memory, we load and validate its candidates by calling Procedure VALIDATEJOISTS. Otherwise, if the size of the chunk exceeds the memory limit, we need to load it in batches. Procedure VALIDATEJOISTS requires all the simplices that could form a joist with a given simplex $\sigma$ to determine the actual joists containing $\sigma$ (Algorithm 2, line 25). To make sure that we process them in the same batch, we sort the chunk by $idx_1$ via an algorithm for external sorting; then we load and validate the candidate joists of one simplex at a time.

### 3.1 Finding the joists

To determine which sets of simplices could form a joist, we borrow techniques from similarity search and build a pruned inverted index for the simplices. To do so, we associate to each simplex $\sigma$ a set of fingerprints, or *codes*, defined as subsets of vertices of $\sigma$. Since a necessary condition for two $q$-simplices $\sigma_1$ and $\sigma_2$ to be in the same joist is to share all but one vertex, we generate codes of size $|\sigma| - 1$. This way, we ensure that each simplex shares one code with all and only candidate cofaces of a joist. Then, Algorithm 2 dynamically creates an inverted index that maps a code $\xi$ to a set of simplices that contain $\xi$. In the first step, the algorithm uses Procedure FIND-MATCHES to find all the simplices $\tau$ that share an indexed code with the current simplex $\sigma$. In the second step, the algorithm indexes all the codes of $\sigma$. Executing FINDMATCHES before indexing the codes of $\sigma$ ensures that each pair of simplices is compared only once.

Once found all the simplices that share a coface $\xi$ with $\sigma$, Procedure MERGE updates the data structure $CJ$, which tracks candidate joists. To ensure that Procedure VALIDATEJOISTS examines each candidate joist only once, we leverage the natural order of the vertices, and thus add to the set of candidates $CJ[\sigma]$ of a simplex $\sigma$ only the simplices $\tau$ such that the vertex $v$ in $\tau$ but not in $\sigma$ satisfies $v > \sigma[|\sigma|-1]$ (Algorithm 2 lines 16-17). This condition is true only for a single $q$-simplex $[v_0, \ldots, v_q]$ in each joist $[v_0, \ldots, v_{q+1}]$.

Procedure VALIDATEJOISTS extracts subsets of simplices in the candidate set $CJ[\sigma]$ of a simplex $\sigma = [v_0, \ldots, v_q]$ that could form, together with $\sigma$, the joist of a $(q+1)$-simplex $[v_0, \ldots, v_q, w]$. Each subset of $(q+1)$ simplices in a joist $J$ of a $(q+1)$-simplex share a vertex $w$, because each pair of simplices in $J$ have $q$ vertices in common. Therefore, for each vertex $w$ in some simplex in $CJ[\sigma]$ but not in $\sigma$, the algorithm checks if the total number of simplices in $CJ[\sigma]$ that contain $w$ is equal to $q+1$ (line 25). When this condition holds, it adds $w$ to the set of real joists $\mathcal{J}[\sigma]$, and $\tau \setminus \sigma$ to the set of real joists $\mathcal{J}[\tau]$ of all the simplices $\tau$ containing $v$ (lines 26-27).

### 3.2 Computing the simplicial trussness values

We first illustrate the algorithm that performs the complete truss decomposition and then explain how the top-$n$ algorithm differs from it. Algorithm 1 initializes the trussness of each simplex $\sigma$ in $E$ as its upper bound (line 10), which is the number of joists that contain $\sigma$. By exploiting an inverted index, the lower bound $lb[\sigma]$ of the simplicial trussness of a simplex $\sigma$ can be computed in linear time. Therefore, the algorithm stores only the simplicial trussness $tr[\sigma]$ of the simplices such that $tr[\sigma] > lb[\sigma]$. If all the simplices $\sigma$ have lower bound equal to the upper bound, the algorithm has already found their real simplicial trussness. Therefore, it inserts them in the set $S$ of simplices to extend in the next iteration, increases $q$, and finally continues to the next iteration. Otherwise, the simplices need to be processed.

To compute the real simplicial trussness values, we insert the simplices in a dictionary $Q$ (line 16) that allows us to extract them in increasing order of upper bound. When examining a simplex $\sigma$, all the simplices in the same joists of $\sigma$ are visited by generating them on the fly. Since the set $CJ$ stores the vertices $v$ such that $\sigma$ is a coface of the $(q+1)$-simplex $\sigma \cup \{v\}$, we can generate those simplices by adding the vertex $v$ to all the subsets of $\sigma$ of size $|\sigma|-1$ (line 22). If some $\tau$ has trussness greater than that of $\sigma$ (line 23), we remove from $CJ[\tau]$ the vertex that belongs to $\sigma$ but not to $\tau$. If this is the first time that we remove it, the simplicial trussness of $\tau$ is decreased by 1 and $Q$ is updated accordingly (line 25). Finally, if $tr[\sigma] > 0$ we insert $\sigma$ in the set of simplices to extend in the next iteration. At the end of the computation, we remove all the simplices $\sigma$ such that $tr[\sigma] = lb[\sigma]$ from $tr$, so that $tr$ contains only the non-trivial simplicial trussness values.

To find the top-$n$ simplices of size $q$ with maximum simplicial trussness for given parameters $n$ and $q$, we modify two procedures of Algorithm 1. First, we let procedure EXTENDSIMPLICES generate only the simplices of size $q$. Secondly, we replace the while cycle at lines 17-27 with Algorithm 3, whose pseudocode is reported in Appendix A. This algorithm examines the simplices in the dictionary $Q$ in descending order of upper bound, to increase the likelihood that the simplices with largest real simplicial trussness are discovered sooner. Then, it exploit a priority queue $T$ of $(t, \varphi)$ pairs sorted on $t$ to maintain the trussness $t$ of the simplices $\varphi$ for which the real simplicial trussness has been found. This way, it can terminate the examination of a simplex $\sigma$ as soon as at least $n$ simplices have been collected in $T$, and the estimate of the trussness of $\sigma$ is lower than the minimum key in $T$ (line 8). Moreover, the algorithm can terminate the computation as soon as the largest upper bound of the simplices still in $Q$ is lower than the minimum key $T$ (line 4). Finally, the algorithm returns the top $n$ elements in $T$, which correspond to the $n$ simplices in $\mathcal{K}$ with maximum trussness.

Table 2: Characteristics of the real datasets.

|  | DBLP | Enron | Zebra | ETFs |
|---|---|---|---|---|
| Vertices | 1 918 581 | 50 224 | 5375 | 2340 |
| Edges | 7 683 001 | 330 179 | 159 134 | 5543 |
| Triangles | 11 350 197 | 1 431 200 | 3 080 990 | 7130 |
| Maximal Simplices | 2 465 299 | 106 442 | 5250 | 2157 |
| Max Dimension | 17 | 64 | 61 | 71 |
| Min/Max Jaccard | 0.01/0.97 | 0.007/0.95 | 0.013/0.81 | 0.07/0.87 |
| Avg/Median Jaccard | 0.61/0.68 | 0.55/0.60 | 0.55/0.61 | 0.39/0.33 |
| Connected Components | 60773 | 1004 | 313 | 174 |

### 3.3 Complexity

Let $\mathcal{K}$ be a simplicial complex with vertex set $V$ of size $m$. In the worst case, $\mathcal{K}$ is a clique complex, i.e., any subset of $V$ belongs to $\mathcal{K}$. In this case, the number of simplices to examine in Algorithm 1 is equal to the number of proper subsets of $V$ with size greater than 1, which is $2^m - m - 2$. Then, each $q$-simplex shares $q-1$ vertices with every other $q$-simplex, and therefore Algorithm 2 has complexity

$$O\left(\sum_{q=2}^{m-1} \binom{m}{q} \cdot \binom{m}{q}\right) \leq O\left(4^m\right)$$

and the total cost of Algorithm 1 is upper-bounded by $O(4^m)$. When the input contains only binary relationships (i.e. it is a graph), the parameter $q$ takes only value 2, and therefore the summation becomes $\binom{m}{2} * \binom{m}{2}$. In this case, the complexity is upper bounded by $O(m^4)$. Let $n$ be the number of binary relationships in the input, then the complexity can be expressed as $O(n^2)$, because the number of binary relationships in a clique complex is $n = m(m-1)/2$.

## 4 EXPERIMENTAL RESULTS

In this section we (1) show the significance of simplicial truss decomposition when compared to classic graph decomposition, (2) evaluate the performance and scalability of our algorithm STRUD, (3) study the *persistent homology* of a dataset by using its simplicial truss decomposition as filtration, (4) show how the simplicial trussness can be used to measure the *manifoldness* of a dataset, (5) compare the simplicial truss decomposition of real datasets with that of random complexes, and (6) present and discuss particular findings in one of the real datasets.

We implemented STRUD in Python 3.6 and Networkx v2.4. We used Dionysus v2.0.7 to compute persistent homology. The code is available on GitHub.[2] We ran the experiments on a 80-Core (2.00 GHz) Intel(R) Xeon(R) Gold 6138 with 126GB of RAM, Ubuntu 18.04, limiting the available memory to 70GB, and using a single core.

We considered four real-world and two synthetic datasets. Their characteristics are reported in Table 2 and Table 3.

**DBLP** is the *coauth-DBLP* simplicial complex provided by Benson et al. [3], where each simplex represents a publication and its vertices are the corresponding authors.

---

[2]https://github.com/lady-bluecopper/STruD

Table 3: Characteristics of the synthetic datasets.

|  | RFC | SCM |
| --- | --- | --- |
| Vertices | 1000 | 360 |
| Edges | 69 274 | 2324 |
| Triangles | 443 033 | 21 023 |
| Maximal Simplices | 254 820 | 191 |
| Max Dimension | 6 | 50 |
| Min/Max Jaccard | 9.11 e−3/0.64 | 0.23/0.6 |
| Avg/Median Jaccard | 0.35/0.30 | 0.46/0.57 |
| Connected Components | 23 | 3 |

**Enron** contains emails sent from roughly 150 employees of the Enron corporation.[3] We obtain a simplicial complex by representing senders and recipients as vertices, and each email as a simplex.

**Zebra** is a simplicial complex constructed from the genetic data of zebrafishes provided by COXPRESdb [26]. Starting from the gene correlation table generated using the method RC-PS[4], we extract, for each gene, the genes with mutual rank value lower than one tenth of the average value in its neighborhood, and then created a simplicial complex containing the extracted genes. Finally, we retain the simplices with size in the 99th percentile.

**ETFs** (Exchange-Traded Funds)[5] contains general aspects, portfolio indicators, returns, and financial ratios of 2353 ETFs, scraped from the Yahoo Finance website.[6] We use this information as features of the ETFs, compute the Kendall correlation between each pair of ETFs, and found the 99.9-th percentile for each row of the correlation matrix. Then, for each ETF, we create a simplex containing all the ETFs with correlation above the 99.9-th percentile. Finally, we retain the simplices with size in the 99.9-th percentile.

**RFC** is a random flag complex [19], i.e., a simplicial complex whose $q$-simplices correspond to the $(q + 1)$-cliques of a random graph. It is constructed by first generating an Erdös-Rényi random graph on $n$ vertices and edge probability $p$, and then creating a $q$-simplex for each $(q + 1)$-clique in the graph. We use $n = 1000$ and $p = 20 \cdot \log(n)/n$.

**SCM** is a null model for comparison with empirical simplicial complexes [42]. Let the *degree* of a vertex in a complex be the number of maximal simplices that contain it, and the *size* of a simplex be the number of vertices it contains. Then, SCM is the uniform distribution over all the complexes with degree sequence $d$ and size sequence $s$. We sample a complex from this distribution by using the uniform Markov chain Monte Carlo sampler (MCMC),[7] which takes in input only a list of maximal simplices from a real complex (we use the Enron complex for this).

### 4.1 Comparison with graph trusses

Figure 2 proves the richness of the simplicial truss decomposition, when compared to the standard truss decomposition. As we can see, the size of the graph trusses is a convex function, whereas the size of the simplicial trusses is concave. This difference is due to the presence of the higher-order structures that exist in the complex, but are lost when adopting a graph representation of the data.

---
[3]https://www.cs.cmu.edu/~enron
[4]https://coxpresdb.jp/download
[5]https://www.kaggle.com/stefanoleone992/mutual-funds-and-etfs
[6]https://finance.yahoo.com
[7]https://github.com/jg-you/scm

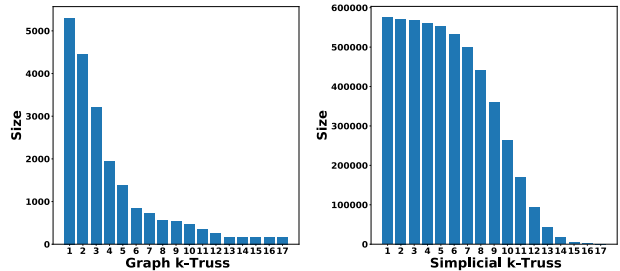

Figure 2: Size of the trusses found in the 1-skeleton of ETFs (left) and size of the simplicial trusses found in ETFs (right).

### 4.2 Scalability

We evaluate the impact of the dataset characteristics on the performance of STruD. Figure 3 (left) shows the total time required to compute the simplicial truss decomposition of the real datasets, varying the maximum size of the simplices to explore. For these experiments, we terminate the computation when the memory size limit is reached, and therefore the running times are not determined by I/O operations. As expected, the time grows exponentially with the maximum size, although more steeply for the Enron and Zebra datasets than for DBLP and ETFs. Even though the size of Enron is smaller than that of DBLP, Enron contains simplices with larger dimension and lower overlap of their vertices, which results in a number of candidate simplices to examine that grows exponentially with the max size. Recall that the number of candidate simplices at each iteration $q$ is upper-bounded by the number of possible subsets of size $q$ of the simplices in the complex. Figure 4 illustrates how the actual number of candidates grows for the DBLP and Enron dataset, together with the running time required to complete each step of the computation. At iteration $q = 5$, the number of candidates in Enron is almost 3 times the number of candidates in DBLP, and hence at iteration $q = 6$, the memory limit is reached. A similar situation happens in the Zebra dataset, where the number of distinct triangles to examine is 3 times the one in Enron.

Instead, Figure 3 (right) shows the time required to find the top-50 simplices of size *Size* with highest simplicial trussness, compared with the running time of the brute force approach (denoted with -B), which first compute the simplicial trussness of the simplices and then retains the top-50. We report only the time required to complete the first step. The chart indicates that the difference between the two approaches is not significant. As illustrated in Figure 4, this is due to the fact that the time required to find (*candidates*) and validate (*validation*) the neighbors dominates the computation, and the two algorithms differ only in how the perform the *trussness* step, i.e., in how they compute the simplicial trussness of the simplices.

These results show that the combinatorial explosion caused by the downward closure property severely affects the performance of any approach that needs to operate on the simplices in a complex.

### 4.3 Persistent homology

Persistent homology [14] is a mathematical tool used in topological data analysis (TDA) to identify the qualitative features of a simplicial complex, and quantify the shape of the underlying data in terms of those features. This is achieved by measuring the lifetime of the topological features through a *filtration*, which is a sequence of nested subcomplexes of the simplicial complex constructed by iteratively adding simplices to an initially empty set, under the condition that a simplex is added only after all its faces. The qualitative features are the topological features with long persistence through the filtration. As such, persistent homology provides a measure of robustness of the features emerging across different scales, and generates an accurate approximation of the underlying data space.

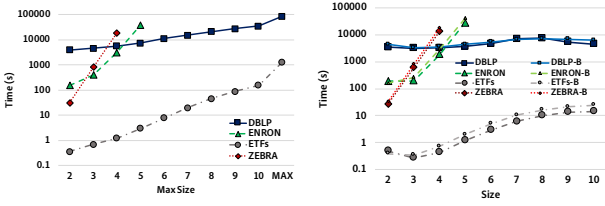

**Figure 3: Running time of STruD to find the simplicial truss decomposition of all the datasets, varying max size of the simplices considered (left); and to find the top-50 simplices with highest simplicial trussness and given size (right).**

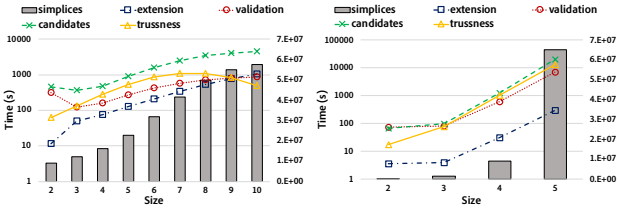

**Figure 4: Running time required by the various steps of STruD in DBLP (left) and Enron (right). The bars indicate the number of simplices of size *Size* that must be examined.**

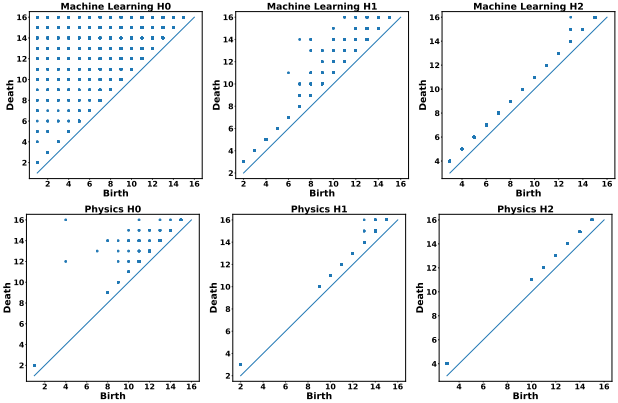

**Figure 5: Persistence diagrams of $H_0$, $H_1$, and $H_2$, for two different subsets of the DBLP dataset.**

Given that the definition of simplicial truss satisfies the containment property, we can use the reverse of the simplicial truss decomposition as a filtration. The persistence homology of a filtration can be visualized via a persistence diagram, which represent each topological feature in the filtration as a point. Each point is called a *persistence pair*, and its coordinates indicate the birth and death time of the feature in the filtration sequence. Roughly speaking, a 0-dimensional feature is a connected component, a 1-dimensional feature is a hole, and a 2-dimensional feature is a void. Figure 5 shows the result of the truss-based persistence homology on different clusters of co-authors in DBLP via persistence diagrams. The left diagrams indicate the 0-dimensional features, the middle ones indicate the 1-dimensional features, while the right ones the 2-dimensional features. Points close to the diagonal are features which are born and immediately die, and thus represent features created by simplices that are not maximal, but contained in a larger simplex. These simplices have trivial simplicial trussness, i.e., simplicial trussness equal to the lower bound. The middle and right parts of Figure 5 (middle and right) show that many of the higher-order simplices in DBLP have trivial simplicial trussness.

Differently, points in the upper-left corner indicate features originated early in the filtration and died at the end of the filtration (or never died). Alive features can correspond to joists of simplices that do not exist in the complex, and can be surrounded by simplices with both high and low simplicial trussness values. The case of high simplicial trussness is particularly interesting, because it corresponds to a dense region of the complex with a hole, i.e., to a group of researchers that frequently collaborated in subgroups but never together. Dead features, instead, can correspond to simplices whose cofaces have simplicial trussness much higher than that of the simplex, meaning that the neighborhood of the simplex is much less dense than those of its cofaces. When looking, for example, for strong collaborations or central vertices in DBLP, one may want to concentrate on these particular cofaces. In contrast, points in the upper-right corner indicate features originated towards the end of the filtration. Since the simplicial trussness is inversely proportional to the time of birth, these points can indicate papers written mostly by small group of authors which do not often cross-collaborate.

Another interesting case is illustrated in the upper-left corner of Figure 5. Given that the simplicial trussness of a simplex is lower bounded by the size of the largest simplex containing it, when the complex contains simplices with heterogeneous sizes, persistence pairs can be found all over the persistence diagram.

Finally, by looking at the persistence diagrams of different datasets, we can compare the dynamics that characterize them. In the case presented in Figure 5, similar diagrams indicate that authors in different fields of research collaborate in a similar way, while different diagrams suggest different underlying mechanisms. The structure and patterns of scientific collaboration have been an object of study for many research communities [13]. Among them, Patania et al. [27] use persistence homology to characterize the patterns of collaboration in the arXiv data.

### 4.4 Manifoldness

It has been shown that the topology and the geometry of a network affect its dynamics [4], and hence play an important role in understanding the organization of the brain [5], and in defining routing protocols [20], among others. A well-studied topological object is the manifold. A *simplicial manifold* is a simplicial complex for which the geometric realization is homeomorphic to a topological manifold, i.e., a space where the neighborhood of each point is homeomorphic to $\mathbb{R}^n$ for some integer $n$. A simplicial manifold $\mathcal{M}$ of dimension $d$ is a growing simplicial complex generated by gluing $d$-simplices along their cofaces [24]. Let $n_\delta$ be the number of $d$-simplices of which $\delta$ is a coface minus 1. In the first step, $\mathcal{M}$ consists in a single $d$-simplex. At any subsequent step $t$, an additional $d$-simplex is attached to a coface $\delta$ in $\mathcal{M}$ with probability:

$$P_\delta = \frac{1 - n_\delta}{\sum_{\delta' \in \mathcal{M}} 1 - n_{\delta'}} \quad.$$

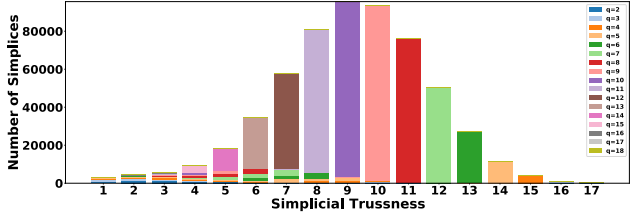

Figure 6: Simplicial Trussness of ETFs.

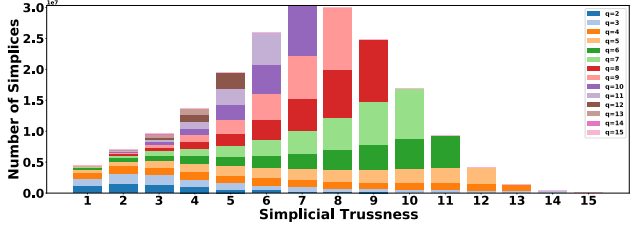

Figure 7: Simplicial Trussness of DBLP.

It follows that the number of vertices in $\mathcal{M}$ grows at each step by one, and hence the total number of vertices $N$ is equal to $N = s + d$, where $s$ is the number of $d$-simplices in $\mathcal{M}$. Given parameters $d$ and $s$, we generated several manifolds of dimension $d$ and size $s$ and performed their simplicial truss decomposition. Experimental results showed that in a manifold of dimension $d$, each $q$-simplex has simplicial trussness equal to $d - q$. This behavior arises from how the simplicial manifolds are generated. By construction, a new vertex is added at each step, meaning that the new vertex $v$ in the last $d$-simplex added is included in $d$ different joists, and therefore its simplicial trussness is equal to $d$. For the same reason, all the 1-simplices incident to $v$ have simplicial trussness $d - 1$, and so on. By cascade, all the neighbors of $v$ have simplicial trussness $d$, all the neighbors of its incident 1-simplices have trussness $d - 1$, and so on. As a consequence, simplicial trussness can quantify how much a dataset deviates from a manifold: the more diverse the simplicial trussness values of simplices of the same size are, the more the simplicial complex is different from a manifold. For example, by looking at Figure 6 and Figure 7 we can conclude that ETFs is more similar to a manifold than DBLP, because most of the simplices of the same size have the same simplicial trussness value.

### 4.5 Truss decomposition of random complexes

Random models are a useful tool to prove the statistical significance of the findings of some analysis on real data. We show that our definition of simplicial trussness is informative and non-trivial by comparing the simplicial truss decomposition of a real complex with that of random complexes.

We generate two random complexes, RFC and SCM, and compute their simplicial truss decomposition. Figure 8 illustrates the size of the simplicial trusses found in Enron (a), RFC (b), and SCM (c), while the simplicial trussness of their simplices is reported in Figure 9 (in

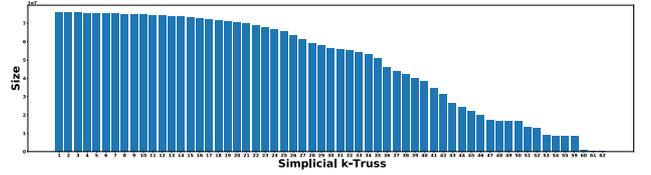

a) Enron

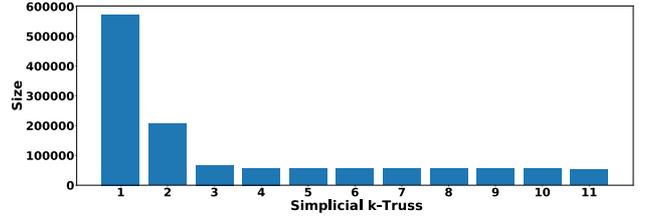

b) RFC

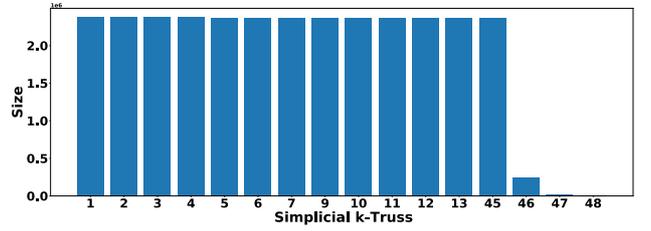

c) SCM

Figure 8: Simplicial trusses of Enron, RFC, and SCM.

Appendix). As the majority of the simplices in SCM have simplicial trussness 45, in Figure 8 (c) we can see that the size of the $k$-trusses with $k \leq 45$ is almost the same. This behavior is due to the presence of a large simplex of size 50 that contains a large portion of the vertices, and thus determines the simplicial trussness of most of the simplices in the complex. In contrast, social networks such as Enron usually follow power-law degree distributions, so that most of the vertices have few connections, and few vertices are highly-connected. As a consequence, the simplicial trussness values of the simplices in Enron are more heterogeneous (Figure 8 (a)) and the convexity of the function in the chart more pronounced. Similarly, RFC does not follow a power-law distribution, having mainly size-5 and size-3 simplices, and therefore, the simplicial trussness values of its simplices are smaller than those of the Enron simplices (Figure 8 (b)). Additionally, Figure 9 (b) shows that most of the higher-order structures have simplicial trussness 1 or 2, hence leading to a $k$-truss size chart that resembles those obtained in the standard truss decomposition. This situation happens when the larger simplices do not participate in many joists, and can be explained by the *simplicial closure phenomenon* [27]: in social networks, three vertices connected in pairs are more likely to form a triangle. Since this principle does not generally hold for random networks, most of the joists are joists of simplices that do not exist in the complex, and thus the simplicial trussness values are low.

The same conclusions can be drawn by looking at Table 4, which summarizes the topological features of Enron, RFC, and SCM, in terms of Betti numbers $\beta_i$ (i.e., structural holes of dimension $i$), percentage of open joists to total joists, percentage of open triangles to total open joists, and percentage of simplices with non-trivial simplicial trussness. Here, we use the term *open* to indicate joists of simplices that do not exist in the complex. As we can see, in RFC 33.5% of the joists are open (the number grows to 50% for SCM) compared to the 0.46% of Enron.

Table 4: Betti numbers $\beta_i$ of Enron, RFC, and SCM; percentage of open joists; percentage of open triangles; and percentage of simplices with non-trivial simplicial trussness.

| | $\beta_0$ | $\beta_1$ | $\beta_2$ | Open Joists (%) | Open Triangles (%) | Non-trivial $tr$ (%) |
|---|---|---|---|---|---|---|
| **Enron** | 215 | 34685 | 28725 | 0.46 | 89.2 | 0.12 |
| **RFC** | 1 | 8 | 275213 | 33.5 | 11.9 | 12.2 |
| **SCM** | 1 | 151 | 79 | 50 | 0.89 | 5e-5 |

Finally, we note that most of the simplices in SCM have trivial simplicial trussness (99.99%), which means that their simplicial trussness value is only determined by the largest simplex that contains them. This number is quite similar to that of Enron (99.87%) due to the simplicial closure phenomenon, but goes down to 87.72 for RFC. Arguably, the simplicial truss decomposition of SCM behaves similarly to that of Enron is because the SCM random complex is generated by using Enron as input.

### 4.6 Analysis of the top simplices

By analyzing the experimental results, it turns out that a simplex $\sigma$ achieves a high simplicial trussness in one of the following two cases: **(i)** $\sigma$ belongs to many overlapping joists, or **(ii)** $\sigma$ is contained in a very large simplex $\tau$. In a collaboration network such as DBLP, these two cases can correspond, respectively, to authors that have often collaborated with each others, and to a paper written by a large group of researchers. When looking for interesting structures in the network, the first situation may be of greater interest than the second one, as it indicates that stronger and more persistent relationships exist between the authors. By storing only the simplices with non-trivial simplicial trussness, we can immediately detect the most interesting simplices in the dataset. Indeed, we note that in the second case the simplicial trussness of $\sigma$ is determined by the size of $\tau$, i.e., it is equal to the lower bound. On the contrary, in the first case, the simplicial trussness of $\sigma$ is non-trivial, and moreover, it is more likely to be much larger than that of the larger simplices $\tau$ containing it. As an example, by looking at the simplices with non-trivial simplicial trussness in DBLP, we can identify the triangle (Marek Tutaj, Howard J. Jacob, Weisong Liu) with simplicial trussness 11, determined by several papers co-written by them. On the other hand, by looking at the top simplices with maximal simplicial trussness, we find that the triangle (Pablo Losada, Charo Gil, Maria C. Viegas) has simplicial trussness 14. However, the latter simplex may be less interesting than the first one, because its simplicial trussness is completely determined by a single paper published in 2011 co-written by 17 authors.

Finally, other interesting structures to analyze are the structural holes, i.e., joists of simplices that do not exist in the complex, surrounded by dense structures, because, due to the simplicial closure phenomenon, they indicate structures that could be filled in the future. This kind of joists is characterized by the presence of higher-order structures with high simplicial trussness around. For example, let consider the joist of the triangle (Liwei Wang, Harold R. Solbrig, Cui Tao). Liwei Wang wrote 3 papers together with Cui Tao, one of which has many authors and thus contributes significantly to the simplicial trussness of the edge (Liwei Wang, Cui Tao). On the other hand, 15 papers contribute to the simplicial trussness of the edge (Cui Tao, Harold R. Solbrig). Finally, one paper with many authors determines the simplicial trussness of the edge (Liwei Wang, Harold R. Solbrig). Even though, Liwei Wang and Harold R. Solbrig work in the same research center and Harold R. Solbrig has an extensive collaboration with Cui Tao, the three of them never collaborated together. However, the high values of simplicial trussness of the edges may be a sign of future collaboration.

## 5 RELATED WORK

**Truss Decomposition.** Truss mining has garnered attention in the data mining community as it represents a cohesive, relaxed version of clique mining that can be computed very efficiently [40]. Its definition is based on the notion of triangles, which have always been considered fundamental building blocks of a network, especially social ones [36]. For this reason, $k$-trusses have been successfully used to detect communities in social networks [21] and to identify target vertices for viral marketing [23], among other applications. Truss decomposition, which is the task of detecting all the non-empty $k$-trusses of a graph, has been studied for probabilistic graphs [15], large graphs [6, 18], bipartite graphs [32] and dynamic graphs [29]. In addition, Sariyuce et al. [33] have proposed a generalization of the $k$-truss decomposition that finds a hierarchy of dense structures in a simple graph. However, computing the truss decomposition of a simplicial complex has not been considered thus far.

**Frequent Itemset Mining.** If we represent a simplicial complex and its simplices as a family of sets, then the simplicial complex can be seen as a transactional database, and the simplices as transactions. This way, the task of finding the top-$n$ simplices with largest trussness resembles that of frequent itemset mining, which requires finding all the subsets of items that appear frequently in a transactional database [1, 17, 25]. Here, the support of an itemset is the number of transaction in the database that contain all the elements in the itemset. Similarly to our case, this definition satisfies the a-priori property, and thus the search space can be efficiently explored in the same bottom-up fashion used in our work. However, an algorithm designed to solve frequent itemset mining cannot be directly applied to solve simplicial truss decomposition: the trussness of a simplex depends on the size of the simplices that contain it and the trussness of the simplices whose vertex set intersect with that of the simplex. Conversely, the support of an itemset depends on the number of transaction that contain it, rather than their size.

**Simplicial Complex Analysis.** Graphs are not always the most suitable data structure to encode relationships between real-world actors. For example, in an email network we cannot distinguish a message with multiple recipients from several messages with a single recipient. Similarly, in a co-authorship network, we cannot distinguish a paper written by a group of authors from several papers written pairs of authors. Simplicial complexes have been used by the data mining community to better capture these higher-order relationships and solve several interesting problems [31]. By observing that the interactions between subsets of a group of users in a social network increase the likelihood that the members of the group will be pairwise connected in the future, Benson et al. [3]

tackled link prediction via simplicial closure. Similarly, Eswaran et al. [12] addressed the semi-supervised learning task of label propagation in partially-labeled graphs. Iacopini et al. [16] adopted simplicial complexes to describe social contagion and diffusion phenomena, Horak et al. [14] to characterize networks by means of their topological features, Serrano and Gómez [34] to define centrality measures, and finally substantial work has been devoted to study the brain's functional and structural organization [22, 28].

## 6 CONCLUSIONS

We introduced the problem of truss decomposition in simplicial complexes, which generalizes the standard truss decomposition on graphs. We showed that our definition of $k$-truss in a complex satisfies the uniqueness, containment, and a-priori property, which allows the development of bottom-up solutions that examine simplices of increasing dimension. Moreover, we identified a convenient lower bound to the simplicial trussness of the simplices that let us reduce the size of the output significantly; and an upper bound that gives an ordering of the simplices to use for the computation of the real values of simplicial trussness. Borrowing ideas from similarity search, we designed STruD, a memory-aware algorithm that can efficiently compute the simplicial truss decomposition of a complex. In addition, we presented a version of STruD that extract the top-$n$ simplices with maximum trussness and given size. Our experimental evaluation has proven **(i)** the richness of the simplicial trusses when compared to the standard ones, and **(ii)** the scalability of STruD; and has shown **(iii)** a topological and **(iv)** a geometrical interpretations of the simplicial trussness (as filtration and as a measure of manifoldness).

**Acknowledgments.** The authors acknowledge support from Intesa Sanpaolo Innovation Center. The funders had no role in study design, data collection and analysis, decision to publish, or preparation of the manuscript.

## REFERENCES

[1] Rakesh Agrawal and Ramakrishnan Srikant. 1994. Fast Algorithms for Mining Association Rules in Large Databases. In *VLDB*.
[2] Ron Atkin. 1974. *Mathematical structure in human affairs*. Heinemann Educational Publishers.
[3] Austin R Benson, Rediet Abebe, Michael T Schaub, Ali Jadbabaie, and Jon Kleinberg. 2018. Simplicial closure and higher-order link prediction. *PNAS* 115, 48 (2018), E11221–E11230.
[4] Ginestra Bianconi and Christoph Rahmede. 2016. Network geometry with flavor: from complexity to quantum geometry. *Physical Review E* 93, 3 (2016), 032315.
[5] Ed Bullmore and Olaf Sporns. 2009. Complex brain networks: graph theoretical analysis of structural and functional systems. *Nature Reviews Neuroscience* 10, 3 (2009), 186–198.
[6] Pei-Ling Chen, Chung-Kuang Chou, and Ming-Syan Chen. 2014. Distributed algorithms for k-truss decomposition. In *Big Data*. 471–480.
[7] Jonathan Cohen. 2008. *Trusses: Cohesive subgraphs for social network analysis*. Technical Report.
[8] Yuri Dabaghian, Facundo Mémoli, Loren Frank, and Gunnar Carlsson. 2012. A topological paradigm for hippocampal spatial map formation using persistent homology. *PLoS Computational Biology* 8, 8 (2012), e1002581.
[9] V. De Silva and R. Ghrist. 2007. Homological sensor networks. *Notices of the American mathematical society* 54 (2007). Issue 1.
[10] W Dörfler and DA Waller. 1980. A category-theoretical approach to hypergraphs. *Archiv der Mathematik* 34, 1 (1980), 185–192.
[11] Ernesto Estrada and Grant J Ross. 2018. Centralities in simplicial complexes. Applications to protein interaction networks. *Journal of Theoretical Biology* 438 (2018), 46–60.
[12] Dhivya Eswaran, Srijan Kumar, and Christos Faloutsos. 2020. Higher-Order Label Homogeneity and Spreading in Graphs. In *The Web Conference 2020*. 2493–2499.
[13] Wolfgang Glänzel and András Schubert. 2004. Analysing scientific networks through co-authorship. In *Handbook of quantitative science and technology research*. 257–276.
[14] Danijela Horak, Slobodan Maletić, and Milan Rajković. 2009. Persistent homology of complex networks. *Journal of Statistical Mechanics: Theory and Experiment* 2009, 03 (2009), P03034.
[15] Xin Huang, Wei Lu, and Laks VS Lakshmanan. 2016. Truss decomposition of probabilistic graphs: Semantics and algorithms. In *ICDM*. 77–90.
[16] Iacopo Iacopini, Giovanni Petri, Alain Barrat, and Vito Latora. 2019. Simplicial models of social contagion. *Nature Communications* 10, 1 (2019), 1–9.
[17] Ruoming Jin and G. Agrawal. 2005. An algorithm for in-core frequent itemset mining on streaming data. In *ICDM*.
[18] Humayun Kabir and Kamesh Madduri. 2017. Shared-memory graph truss decomposition. In *HiPC*. 13–22.
[19] Matthew Kahle. 2009. Topology of random clique complexes. *Discrete Mathematics* 309, 6 (2009), 1658–1671.
[20] Robert Kleinberg. 2007. Geographic routing using hyperbolic space. In *INFOCOM*. 1902–1909.
[21] Qing Liu, Minjun Zhao, Xin Huang, Jianliang Xu, and Yunjun Gao. 2020. Truss-based community search over large directed graphs. In *ICDM*. 2183–2197.
[22] Louis-David Lord, Paul Expert, Henrique M. Fernandes, Giovanni Petri, Tim J. Van Hartevelt, Francesco Vaccarino, Gustavo Deco, Federico Turkheimer, and Morten L. Kringelbach. 2016. Insights into Brain Architectures from the Homological Scaffolds of Functional Connectivity Networks. *Frontiers in Systems Neuroscience* 10 (2016), 85.
[23] Fragkiskos D Malliaros, Maria-Evgenia G Rossi, and Michalis Vazirgiannis. 2016. Locating influential nodes in complex networks. *Scientific reports* 6 (2016), 19307.
[24] Ana P Millán, Joaquín J Torres, and Ginestra Bianconi. 2018. Complex network geometry and frustrated synchronization. *Scientific Reports* 8, 1 (2018), 1–10.
[25] S. Moens, E. Aksehirli, and B. Goethals. 2013. Frequent Itemset Mining for Big Data. In *BigData*. 111–118.
[26] Takeshi Obayashi, Yuki Kagaya, Yuichi Aoki, Shu Tadaka, and Kengo Kinoshita. 2018. COXPRESdb v7: a gene coexpression database for 11 animal species supported by 23 coexpression platforms for technical evaluation and evolutionary inference. *Nucleic Acids Research* 47, D1 (2018), D55–D62.
[27] Alice Patania, Giovanni Petri, and Francesco Vaccarino. 2017. The shape of collaborations. *EPJ Data Science* 6, 1 (2017), 18.
[28] Giovanni Petri, Paul Expert, Federico Turkheimer, Robin Carhart-Harris, David Nutt, Peter J Hellyer, and Francesco Vaccarino. 2014. Homological scaffolds of brain functional networks. *Journal of The Royal Society Interface* 11, 101 (2014).
[29] Venkata Rohit Jakkula and George Karypis. 2019. Streaming and Batch Algorithms for Truss Decomposition. *arXiv preprint arXiv:1908.10550* (2019).
[30] Manish Saggar, Olaf Sporns, Javier Gonzalez-Castillo, Peter A Bandettini, Gunnar Carlsson, Gary Glover, and Allan L Reiss. 2018. Towards a new approach to reveal dynamical organization of the brain using topological data analysis. *Nature Communications* 9, 1 (2018), 1–14.
[31] Vsevolod Salnikov, Daniele Cassese, and Renaud Lambiotte. 2018. Simplicial complexes and complex systems. *European Journal of Physics* 40, 1 (2018).
[32] Ahmet Erdem Sarıyüce and Ali Pinar. 2018. Peeling bipartite networks for dense subgraph discovery. In *Proceedings of the Eleventh ACM International Conference on Web Search and Data Mining*. 504–512.
[33] Ahmet Erdem Sariyuce, C Seshadhri, Ali Pinar, and Umit V Catalyurek. 2015. Finding the hierarchy of dense subgraphs using nucleus decompositions. In *Proceedings of the 24th International Conference on World Wide Web*. 927–937.
[34] Daniel Hernández Serrano and Darío Sánchez Gómez. 2019. Centrality measures in simplicial complexes: applications of TDA to Network Science. *arXiv preprint arXiv:1908.02967* (2019).
[35] L. M. Seversky, S. Davis, and M. Berger. 2016. On time-series topological data analysis: new data and opportunities. In *CVPR Workshops*. 59–67.
[36] Georg Simmel. 1950. *The Sociology of Georg Simmel*. Vol. 92892. Simon and Schuster.
[37] David I Spivak. 2009. Higher-dimensional models of networks. *arXiv preprint arXiv:0909.4314* (2009).
[38] Leo Torres, Ann S Blevins, Danielle S Bassett, and Tina Eliassi-Rad. 2020. The why, how, and when of representations for complex systems. *arXiv preprint arXiv:2006.02870* (2020).
[39] Jeremy D. Turiela, Paolo Barucca, and Tomaso Astea. 2020. Simplicial persistence of financial markets: filtering, generative processes and portfolio risk. *arXiv preprint arXiv:2009.08794* (2020).
[40] Jia Wang and James Cheng. 2012. Truss decomposition in massive networks. *arXiv preprint arXiv:1205.6693* (2012).
[41] Kelin Xia and Guo-Wei Wei. 2014. Persistent homology analysis of protein structure, flexibility, and folding. *Int J Numer Meth Bio* 30, 8 (2014), 814–844.
[42] Jean-Gabriel Young, Giovanni Petri, Francesco Vaccarino, and Alice Patania. 2017. Construction of and efficient sampling from the simplicial configuration model. *Physical Review E* 96, 3 (2017).

## A TOP-N SIMPLICES OF SIZE Q

Algorithm 3 reports the pseudocode of the method for finding the top-$n$ simplices with size $q$ and maximal simplicial trussness, discussed in Section 3.2.

**Algorithm 3** Top-$n$ Simplices of size $q$

**Require:** Simplicial complex $\mathcal{K}$; Num simplices $n$; Size $q$
**Ensure:** Top-$n$ simplices of size $q$ with maximum trussness
1: $T \leftarrow$ priority queue of $(t, \varphi)$ pairs sorted on $t$
2: **while** $Q \neq \emptyset$ **do**
3:     $M \leftarrow$ simplices with max trussness upper bound $tr_M$ in $Q$
4:     **if** $|T| \geq n$ **and** $tr_M \leq T.min\_key()$ **then**
5:        **break**
6:     **for** $\sigma \in M$ **do**
7:        **for** $v \in CJ[\sigma]$ **do**
8:           **if** $|T| \geq n$ **and** $tr[\sigma] \leq T.min\_key()$ **then**
9:              **break**
10:           **for** subset $\xi$ of $\sigma$ of size $|\sigma| - 1$ **do**
11:              $\tau \leftarrow \xi \cup \{v\}$
12:              **if** $tr[\tau] < tr[\sigma]$ **then**
13:                 $CJ[\sigma] \leftarrow CJ[\sigma] \setminus \{v\}$
14:                 $tr[\sigma] \leftarrow |CJ[\sigma]|$
15:                 **break**
16:     remove $\sigma$ from $Q$
17:     insert $(tr[\sigma], \sigma)$ in $T$
18: **return** $T.top(n)$

## B PROOFS

We next report the proof of Property 1 (uniqueness): i.e., the $k$-truss of a simplicial complex $\mathcal{K}$ is unique.

PROOF. Assume, by absurd, that $T_k^1$ and $T_k^2$ are two distinct $k$-trusses of $\mathcal{K}$. Let $T_k^3 = T_k^1 \cup T_k^2$. For each $\sigma \in T_k^3$, $tr(\sigma) \geq k$ because each $\sigma$ in either $T_k^1$ or $T_k^2$ has trussness $\geq k$ by definition. Since the size of $T_k^3$ is not lower than that of $T_k^1$ and $T_k^2$, each $q$-simplex in $T_k^3$ is involved in not less than $k$ $(q+1)$-ary relationships, meaning that the number of joists to which $\sigma^{(q)}$ belongs is not lower than $k$. Therefore $T_k^3$ is a $k$-truss of $\mathcal{K}$ larger than $T_k^1$ and $T_k^2$, which is a contraction as $T_k^1$ and $T_k^2$ are maximal. □

We next report the proof of Property 2 (containment): i.e., the $(k+1)$-truss of a simplicial complex $\mathcal{K}$ is a subset of its $k$-truss.

PROOF. Each $\sigma \in T_{k+1}$ has trussness $tr(\sigma) \geq k+1$, meaning that it belongs to at least $k+1$ joists. Therefore, it satisfies the condition to belong to $T_k$. As a result, it holds that $T_{k+1} \subset T_k$. □

We next report the proof of Property 3 (a-priori property): i.e., a $q$-simplex $\sigma^{(q)}$ that is a face of a $(q+1)$-simplex $\sigma^{(q+1)}$ has trussness $tr_{\mathcal{K}}(\sigma^{(q)}) \geq tr_{\mathcal{K}}(\sigma^{(q+1)})$.

PROOF. Let assume there exists a $(q+1)$-simplex $\sigma^{(q+1)}$ that contains $\sigma^{(q)}$ and has trussness $k+1$ with $k = tr_{\mathcal{K}}(\sigma^{(q)})$. Then, $T_{k+1}$ contains $k+1$ joists $J_i$, $i = 1, \ldots, k+1$ of $(q+2)$-simplices $\sigma^{(q+2)}{}_i$, each of which contains $(q+3)$ $(q+1)$-simplices with trussness $k + 1$. Each $\sigma^{(q+2)}{}_i$ contains the same vertices of $\sigma^{(q+1)}$ plus an additional vertex $v_i$. Since each pair of cofaces of a $q$-simplex share exactly $q - 1$ vertices, for each $J_i$ there exists a pair of simplices that share the coface $\sigma^{(q)}$, for a total of $k + 2$ $(q+1)$-simplices. Similarly, there exist a set of $k+2$ $(q+1)$-simplices that contain each other coface of $\sigma^{(q+1)}$, meaning that $\sigma^{(q)}$ and all the other $q$-simplices in $\sigma^{(q+1)}$ appear in $k + 2$ joists. Therefore, the simplicial trussness of $\sigma^{(q)}$ is $k+2$. We reached a contradiction, as we assumed $k = tr_{\mathcal{K}}(\sigma^{(q)})$. □

## C ADDITIONAL EXPERIMENTS

Figure 9 illustrates the number of simplices per simplicial trussness value, for Enron (a), RFC (b), and SCM (c). These charts supplement those presented in Section 4.5.

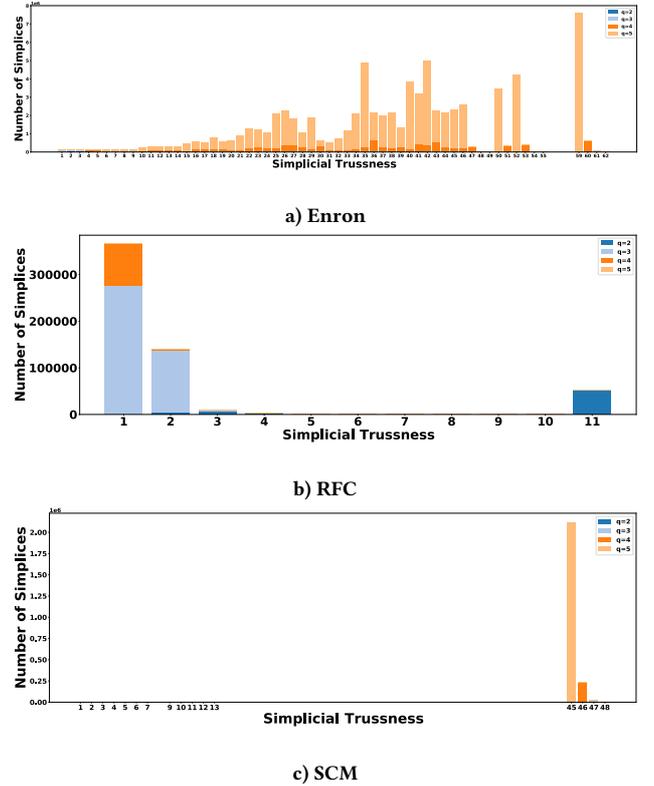

a) Enron

b) RFC

c) SCM

Figure 9: Simplicial trussness in Enron, RFC, and SCM.